\documentclass[prl,twocolumn,aps,amssymb,showpacs,superscriptaddress,nofootinbib]{revtex4}
\usepackage{graphicx}

\begin{document}
%

\title{Structure formation and the origin of dark energy}

\author{Golam Mortuza Hossain}
\email{hossain@gravity.psu.edu}
\affiliation{
Institute for Gravitation and the Cosmos,
The Pennsylvania State University,\\
104 Davey Lab, University Park, PA 16802, USA}

\begin{abstract}

Cosmological constant {\em a.k.a.} dark energy problem is considered
to be one major challenge in modern cosmology. Here we present a model
where large scale structure formation causes spatially-flat FRW
universe to fragment into numerous `FRW islands' surrounded by
vacuum. We show that this mechanism can explain the origin of dark
energy as well as the late time cosmic acceleration. This explanation
of dark energy does {\em not} require any exotic matter source nor an
extremely {\em fine-tuned} cosmological constant. This explanation is
given within classical general relativity and relies on the fact that
our universe has been undergoing structure formation since its recent
past.

\end{abstract}

\preprint{IGC--07/9--3}

\maketitle


Several recent experimental observations \cite{cmb} seem to strongly
suggest that we live in the universe whose energy budget is dominated by
contribution from a mysterious source which is otherwise invisible or
missing in direct observations. This energy component is generally
referred as {\em dark energy}. Furthermore, the dark energy component
appears to have {\em negative} pressure. Result from supernova
observations \cite{supernova} that universe is undergoing a recent
{\em acceleration} seems to confirm such peculiar behavior of dark
energy.

Many attempts have been made in literature to understand the origin of
dark energy (see \cite{ccreview} for some reviews). Arguably the most
economical one is to introduce a non-zero {\em cosmological constant}
in Einstein equation. However, there are several conceptual
difficulties associated with it. Firstly, experimentally required
value of cosmological constant turns out to be $\sim 10^{-123}$ in
Planck unit.  Such an extremely {\em fine-tuned} value of cosmological
constant appears to defy any hope of possible explanation from a
fundamental theory. Second conceptual problem is the so-called {\em
cosmic coincidence} problem: why does energy density due to
cosmological constant which remains unchanged, become comparable to
changing matter energy density only at {\em current} epoch?
%

There have been other attempts where it is anticipated that late time
acceleration could be due to a dynamical scalar field that evolves in
a suitably engineered potential \cite{dynade}, due to super-horizon
perturbations \cite{inhomo}, from averaging of LTB models \cite{ltb}
or due to {\em back-reaction} effects of inhomogeneous structures in
the universe \cite{backreaction,rasanen:curvature,wiltshire}. In particular,
using Buchert equations one can get acceleration by incorporating
inhomogeneous back-reaction. However, same back-reaction term
contributes {\em negatively} \cite{rasanen:curvature} to energy
density thus fails to explain missing energy unless one also
introduces strong {\em negative} spatial curvature. In brief, none of
the current attempts can explain the origin of dark energy
satisfactorily.

\begin{figure}
\includegraphics[width=4cm,height=4cm]{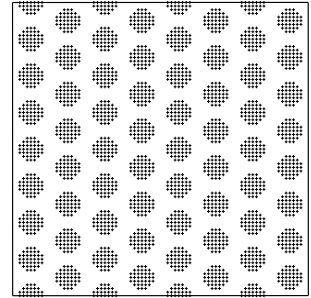}
\caption{A simple illustration of late time universe where structure
formation causes FRW universe to fragment into numerous `FRW islands'.
Each FRW island (shaded region) is surrounded by vacuum metric. 
\label{fig1}
}
\end{figure}

According to standard model of cosmology, continued expansion of the
universe causes radiation to eventually decouple from matter.
Subsequently, largely homogeneous matter distribution with small
inhomogeneity starts collapsing to give rise current large scale
structures. Initial phase of structure formation can be described by
linear perturbation theory around homogeneous background. However, in
later phase when structure formation enters non-linear regime such
descriptions are insufficient. In this era, matter distribution
consists of numerous dense regions surrounded by relatively rarer
regions. 
For simplicity here we make {\em sharp-boundary} approximation for the
matter distribution during later period of structure formation such
that matter is contained within spherical, homogeneous regions
surrounded by empty space.  Boundary of such spherical regions,
however, needs to be continuously refined as on-going structure
formation continues to cause dense regions to become denser.  In other
words, in this model structure formation along with sharp-boundary
approximation causes FRW universe to fragment into numerous `FRW
islands' with shrinking boundary that are surrounded by vacuum (see
Fig.\ref{fig1}).
We will show that this mechanism can explain the origin of dark energy
as well as the late time cosmic acceleration. 
%
%

\begin{figure}
\includegraphics[width=6cm,height=6cm]{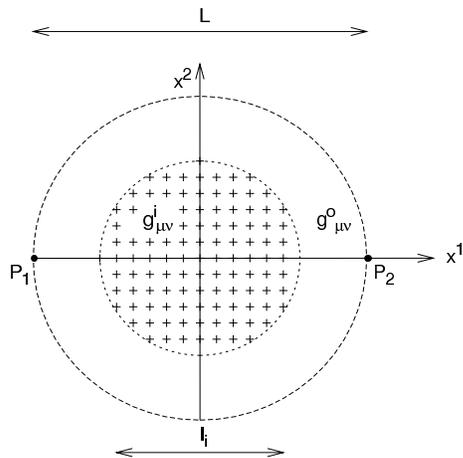}
\caption{A spherical patch of universe of coordinate diameter $L$
which was homogeneous at the beginning of structure formation. In a
later period the matter distribution, by means of sharp-boundary
approximation, is contained within a spherical region (shaded) of
coordinate diameter $l_i$. 
\label{fig2}
} 
\end{figure}

In the standard model of cosmology, spacetime describing our universe
is assumed to be foliated by homogeneous and isotropic spatial
hyper-surfaces parametrized by a global time. We imagine such an
observer who treats the spatial hyper-surfaces as homogeneous and
isotropic during {\em entire} evolution of universe and accordingly
measures distance using an average spatially-flat
Friedman-Robertson-Walker (FRW) metric
\begin{equation}\label{FRWMetric} 
ds^2 = - dt^2 + a^2(t)d{\bf x}^2 ~, 
\end{equation} 
where $a(t)$ is the {\em scale factor}. We refer this observer as
observer A. We consider another observer, say observer B, 
who uses same time parametrization of the spatial hyper-surfaces as
observer A,
but is careful to consider the effects of structure formation. In
particular, to measure distance during structure formation observer B
uses flat FRW metric $g_{\mu\nu}^i = diag(-1, a_i^2, a_i^2, a_i^2)$
inside the spherical regions containing homogeneous matter
distribution otherwise uses vacuum metric $g_{\mu\nu}^o$.
To illustrate this, let's consider a spherical patch of universe of
coordinate (also co-moving for observer A) diameter $L$ which is
homogeneous at the beginning of structure formation (see
Fig.\ref{fig2}). The patch then begins to undergo structure formation
such that at a given time collapsing matter distribution, by means
of sharp-boundary approximation, is contained within a spherical
region of coordinate diameter $l_i(t)$ with $0<l_i\le L$.  We {\em
assume} that the vacuum metric $g_{\mu\nu}^o$ describing the
spherically symmetric empty space created due to structure
formation, is {\em static}. 
(This would be the case if we ignore the presence of neighbouring
patches, as then {\em Birkhoff's theorem} would imply spherically
symmetric vacuum metric is static.)
The vacuum metric $g_{\mu\nu}^o$ however is inhomogeneous as can be
seen from {\em continuity} of metric at the boundary of inside region.

To have a precise notion we define {\em average} FRW metric
(\ref{FRWMetric}) such that the {\em proper} distance between the
points $P_1$ and $P_2$ (see Fig.\ref{fig2}) as measured by observer A
is {\em equal} to the proper distance measured by observer B. Without
loss of generality we assume that the points lie on $x^1$-axis. Given
elementary proper distance in a spatial hyper-surface is
$|\sqrt{g_{jk}dx^jdx^k}|$ where $j$, $k$ represent spatial
coordinates, the definition leads to 
\begin{equation}\label{ProperDistanceAB}
a L = a_i l_i+2\int_{l_i/2}^{L/2}\sqrt{g^o_{11}(x^1,x^2,x^3)}dx^1 ~.
\end{equation}
Using the definition, we can also compute relation between expansion
rates of the points as measured by both observers. In particular,
\begin{eqnarray}\label{ExpansionRateAB}
\dot{a}L &=& 
\frac{d}{dt} \int_{-L/2}^{L/2} \sqrt{g_{11}(x^1,x^2,x^3)}dx^1 
\nonumber\\
&=& \dot{a}_i l_i + a_i \dot{l}_i -
\dot{l}_i \sqrt{g^o_{11}(l_i/2,0,0)} = \dot{a}_i l_i ~,
\end{eqnarray}
where over-dot denotes derivative {\em w.r.t.} to time $t$.  In last
line, we have used continuity of metric solution {\em i.e.}
$g^o_{11}(l_i/2,0,0)=g_{11}^i=a_i^2$.  We may note that to have an
expanding average metric, the metric of the inside region must
also be expanding. However, proper volume of the inside region can
still decrease as its coordinate diameter shrinks. 
We now define average energy density $\rho$ for observer A, by
requiring that at any given time total energy contained within the
patch of coordinate diameter $L$ is {\em same} as measured by observer
B. In particular, if observer B measures energy density of inside
region to be $\rho_i$ and of outside region to be zero then 
\begin{equation}\label{RhoRelation}
\rho = \left(\frac{a_i l_i}{a L}\right)^3 \rho_i =: n^3 \rho_i ~. 
\end{equation}
As defined, $n^3$ is the {\em fraction} of total proper volume occupied
by matter distribution. Physically, parameter $n$ is a measure of {\em
amount} of structure formation and satisfies $0<n\le 1$. In particular,
$n=1$ implies there are no underlying structures. This can be seen
from the equation (\ref{ProperDistanceAB}).
%
As we will see, equations (\ref{ExpansionRateAB}) and
(\ref{RhoRelation}) form the backbone of arguments presented here.

Let's imagine that both observers want to confront their respective
Einstein equations with experimental data. Observer A performs
separate observations to measure expansion rate as well as energy
density. However, it turns out that to make a right balance, observer
A needs to postulate an extra {\em invisible} component in Friedmann
equation for the average metric {\em i.e.}
\begin{equation}\label{FriedmannEqnA} 
3\left(\frac{\dot a}{a}\right)^2 
= 8\pi G(\rho + \rho_{DE}) ~, 
\end{equation} 
where $G$ is Newton's
constant and $\rho_{DE}$ denotes {\em dark energy} component. On the
other hand observer B uses standard Friedmann equation for the region
containing homogeneous matter distribution and uses vacuum Einstein
equation for the remaining region. In particular, Friedmann equation
for observer B is
\begin{equation}\label{FriedmannEqnB} 
3\left(\frac{\dot a_i}{a_i}\right)^2 = 8\pi G \rho_i ~.  
\end{equation} 
Equipped with the details of underlying structures {\em i.e.} using
equations (\ref{ExpansionRateAB}), (\ref{RhoRelation}) and Friedmann
equation (\ref{FriedmannEqnB}), observer B can derive Friedmann
equation for average FRW metric (\ref{FRWMetric}), given by
\begin{equation}\label{FriedmannEqnAFromB} 
3\left(\frac{\dot a}{a}\right)^2 = 
8\pi G\left[\rho + \left(\frac{1}{n}-1\right)\rho\right] ~.
\end{equation} 
One may note that right hand side of equation
(\ref{FriedmannEqnAFromB}) has an extra energy density component apart
from average energy density $\rho$. Thus, comparing equations
(\ref{FriedmannEqnA}) and (\ref{FriedmannEqnAFromB}), observer B can
{\em derive} the expression of dark energy that observer A should
perceive 
\begin{equation}\label{DarkEnergyExpr} 
\rho_{DE} = \left(\frac{1}{n}-1\right)\rho ~.  
\end{equation}
In the situation when $n=1$, dark energy component disappears. In
other words, observer A wouldn't have perceived any dark energy
component if there were {\em no} underlying structures in universe.
Since existence of underlying structures requires parameter values to
be $0<n<1$ then dark energy component is necessarily {\em positive}.
This is in {\em contrast} with back-reaction models such as
\cite{backreaction,rasanen:curvature} where back-reaction term
contributes {\em negatively} to energy density. As evident, dark
energy component is comparable to the magnitude of average energy
density $\rho$. So this model can naturally explain {\em cosmic
coincidence} problem. Another crucial property of dark energy
component (\ref{DarkEnergyExpr}) is that it may appear as {\em
constant} even though it is naively proportional to decreasing average
energy density $\rho$. With the beginning of structure formation, the
value of parameter $n$ starts decreasing from unity. So during
structure formation proportionality factor $(1/n-1)$ increases. This
implies that for suitable rate of structure formation dark energy
component (\ref{DarkEnergyExpr}) may appear as constant.

For observer A total energy contained within the patch is given
by $E=V\rho$ where proper volume of the patch $V=(\pi a^3L^3/6)$.
Using the definition of pressure $P=-(\partial E/\partial V)$, one can
derive the {\em conservation} equation for energy density $\rho$ as
\begin{equation}\label{ConservationEqn}
\dot{\rho} = -\left(\frac{\dot V}{V}\right)
\left(\frac{E}{V} -\frac{\partial E}{\partial V}\right)
= -3 H \rho (1+\omega) ~, 
\end{equation}
where $H := ({\dot a}/a)$ is {\em Hubble parameter} and
$\omega:=P/\rho$ is the corresponding equation of state. Analogously,
we can define equation of state $\omega_{DE}$ for dark energy
component such that $\dot{\rho}_{DE} = -3 H \rho_{DE}
(1+\omega_{DE})$. Using equations (\ref{DarkEnergyExpr}) and
(\ref{ConservationEqn}), we can compute equation of state for the dark
energy
\begin{equation}\label{DarkEnergyEoS}
\omega_{DE} = -1+\left[(1+\omega) - \frac{r_n}{3(1-n)}\right]~, 
\end{equation}
where $({\dot n}/n) =: -r_n H$. Parameter $r_n$ is a measure of {\em
rate} of structure formation. 
Dark energy expression (\ref{DarkEnergyExpr}) along with its equation
of state (\ref{DarkEnergyEoS}) can mimic a cosmological constant at
current epoch for suitable values of structure formation parameters
$n$ and $r_n$. In particular, the values of structure formation
parameters such that $r_n= 3(1-n)(1+\omega)$ will lead to
$\omega_{DE}=-1$ which is the equation of state for a cosmological
constant. However, a distinguishing feature between them is that while
equation of state for a cosmological constant remains unchanged, the
equation of state (\ref{DarkEnergyEoS}) varies with time.

To explicitly show that structure formation can lead to an
accelerating phase, it is convenient to compute Raychaudhuri equation.
Taking time-derivative of equation (\ref{FriedmannEqnAFromB}) and then
using conservation equation (\ref{ConservationEqn}), we can derive
Raychaudhuri equation as 
\begin{equation}\label{RaychaudhuriEqn}
3\left(\frac{\ddot a}{a}\right) = 
4\pi G\left[\frac{r_n}{n}-\frac{(1+3\omega)}{n}\right]\rho ~.
\end{equation}
From equation (\ref{RaychaudhuriEqn}) it can be seen that the values
of structure formation parameter such that $r_n > (1+3\omega)$ will
lead to an accelerating phase for observer A.

The modifications to average dynamics of the given patch due to
underlying structures, depend on the values of parameters $n$ and
$r_n$.  However, the modifications do {\em not} depend explicitly on
coordinate diameters $L$ and $l_i$. Thus, if one considers a different
patch but with {\em same} values of parameters $n$ and $r_n$, then one
will get {\em same} Friedmann and Raychaudhuri equations. Given one
can pack $\mathbb R^3$ space very closely using 3-spheres of {\em
arbitrary} diameters hence the modified Friedman equation
(\ref{FriedmannEqnAFromB}) and modified Raychaudhuri equation
(\ref{RaychaudhuriEqn}) can be considered as good approximation of
equations that describe average dynamics of 
the model 
universe with underlying structures. 
Observed contribution from the dark energy component at current epoch
is about $70\%$ of the critical energy density. The equation
(\ref{DarkEnergyExpr}) with the parameter value $n=0.3$ can lead to
such an observed amount of dark energy. If we consider the average
matter to be pressure-less {\em i.e.} $\omega=0$ then $\omega_{DE}=-1$
requires parameter value $r_n=2.1$. 

To characterize the matter distribution of inside region as seen by
observer B, we need to compute relation between average equation of
state $\omega$ and equation of state for inside region $\omega_i:=
P_i/\rho_i$ where pressure $P_i= -(\partial E_i/\partial V_i)$. $E_i$
and $V_i$ are total energy and proper volume of the inside region
respectively. As earlier we can derive the conservation equation for
inside region, given by
\begin{eqnarray}\label{ConservationEqnB}
\dot{\rho}_i &=& -\left(\frac{\dot V_i}{V_i}\right)
\left(\frac{E_i}{V_i} -\frac{\partial E_i}{\partial V_i}\right)
\nonumber\\
&=& -3 \left(\frac{\dot a_i}{a_i}+ \frac{\dot l_i}{l_i}\right) 
\rho_i (1+\omega_i) ~. 
\end{eqnarray}
Coordinate diameter of inside region is time-dependent and it is
reflected in conservation equation (\ref{ConservationEqnB}) with its
explicit dependence on $(\dot l_i/l_i)$. Using the relation between
energy density (\ref{RhoRelation}), their conservation equations
(\ref{ConservationEqn}) and (\ref{ConservationEqnB}), one can compute
relation between the equation of states
\begin{equation}\label{EoSB}
\omega = (1-r_n)\omega_i ~.
\end{equation}
For gravitational collapse to occur with ordinary matter, the
corresponding matter distribution must be pressure-less. So for
observer B, matter distribution of inside region should be
pressure-less {\em i.e.} $\omega_i=0$. On the other hand observer A
finds average matter also to be pressure-less {\em i.e.} $\omega=0$.
The equation (\ref{EoSB}) ensures that physical requirement of
observer B and observed fact for observer A can be consistently met.
We should mention here that to derive equation
(\ref{RaychaudhuriEqn}), one can also use Raychaudhuri equation for
observer B. However, in doing so one should be careful to include the
additional pressure component coming from the shrinking boundary.

Experimental observations seem to also imply that total energy of the
universe has another dark component, the so-called {\em dark matter}
which is pressure-less. From equation (\ref{DarkEnergyEoS}), one may
note that if $\omega=0$ and $r_n=0$ {\em i.e.} if $(-\dot l_i/l_i)=
(1/n -1)H$ then the equation of state mimics a pressure-less energy
component. It is conceivable that some `FRW islands' may have
different values of structure formation parameters leading to such
behavior. However, whether such scenario can explain phenomena
ascribed to the presence of dark matter such as galaxy rotation
curves, remains to be explored.

To summarize, we have argued that the origin of dark energy can be
understood as a consequence of large scale structure formation. This
explanation of dark energy does {\em not} require any exotic matter
source nor a fine-tuned cosmological constant. However, presented
model in its current form has several deficiencies. Firstly, we assume
that structure formation leads to creation of void around each FRW
island. However, we know cosmic microwave background (CMB) photons
fills up entire universe. Thus, even though CMB contribution to
average energy density is negligible during structure formation but
for an accurate description one should consider their presence. In
this model net effects of structure formation on dynamics of average
metric can be summarized by introducing just two characteristic
parameters $n$ and $r_n$. However, the model itself does {\em not}
shed any light on the values of the parameters $n$ and $r_n$. We may
recall that the model is based on sharp-boundary approximation of the
matter distribution which is under-going structure formation. Thus, to
compute relation between the values of the parameters one needs to
perform a detailed simulation of structure formation with a matter
distribution which should be then successively approximated by
sharp-boundary approximation. Finally, we have not addressed the
issue: why is cosmological constant {\em zero} in our universe?

{\em Acknowledgments:} Author thanks Abhay Ashtekar for discussions
and Martin Bojowald for comments on the manuscript. This work was
supported in part by NSF grant PHY0456913.

\end{document}